# Effect of Oxygen and Aluminium Incorporation on Local Structure of GaN Nanowires: Insight from Extended X-ray Absorption Fine Structure Analysis

Santanu Parida,[†,*,#] Madhusmita Sahoo,[†,#] Abharana N,[‡] Raphael M. Tromer,[§,#] Douglas S. Galvao, [§,¶,*] and Sandip Dhara[†,*]

[†] Surface and Nanoscience Division, Indira Gandhi Centre for Atomic Research, Homi Bhabha National Institute, Kalpakkam-603102, India

[‡] Atomic and Molecular Physics Division, Bhabha Atomic Research Centre, Mumbai-400085, India

[§] Applied Physics Department, State University of Campinas, Campinas, SP, 13083-970, Brazil

[¶] Center for Computational Engineering and Sciences, State University of Campinas, Campinas, SP, 13083-970, Brazil

[#] Authors have equal contribution

[*] Corresponding author email: santanuparida026@gmail.com, galvao@ifi.unicamp.br, dhara@igcar.gov.in

**Abstract:** A thorough investigation of local structure, influencing macroscopic properties of the solid is of potential interest.We investigated the local structure of GaN nanowires (NWs) with different native defect concentration synthesized by the chemical vapor deposition technique. Extended X-ray absorption fine structure (EXAFS) analysis and semi-empirical and the density functional theory (DFT) calculations were used to address the effect of dopant incorporation along with other defects on the co-ordination number and bond length values. The decrease of the bond length values along preferential crystal axes in the local tetrahedral structure of GaN emphasizes the preferred lattice site for oxygen doping. The preferential bond length contraction is corroborated by the simulations. We have also studied the impact on the local atomic configuration of GaN NWs with Al incorporation. $Al_xGa_{1-x}N$ NWs are synthesized *via* novel ion beam techniques of ion beam mixing and post-irradiation diffusion process. The change in the local tetrahedral structure of GaN with Al incorporation is investigated by EXAFS analysis. The analysis provides a clear understanding of choosing a suitable process for ternary III-nitride random alloy formation. The local structure study with the EXAFS analysis is corroborated with the observed macroscopic properties studied using Raman spectroscopy.

# 1. INTRODUCTION

III-V nitrides and their ternary alloys attract tremendous attention in the scientific community as well as the optoelectronic industry due to their applications in light-emitting diodes (LEDs) and laser diodes.[1-4] These materials possess high carrier mobility, elevated melting temperature, high thermal conductivity, enormous physical hardness, and extremely large hetero-junction band offsets.[5-7] The ternary alloys of III-V nitrides also offer a unique opportunity to tune the bandgap from ~0.7 to 6.2 eV by varying the atomic percentage of the constituent elements.[8-10] In the recent past 1D nanostructures of III-V nitrides have emerged as ideal candidates for nanoscale devices because of the high surface area to volume ratio with controllable recombination of carriers, enhanced light-matter interaction, insignificant presence of extended defects, and reduced cost and materials consumption. Apart from the concentration of the constituent elements in the alloys, the point defects also play a major role in the optimum performance of an optoelectronic device.[5] Particularly, in the case of III-V nitrides, native point defects play a vital role in controlling the carrier density.[11,12] The commonly found native point defects in III-V nitrides are nitrogen vacancies ($V_N$), doping of oxygen at lattice site of nitrogen ($O_N$),[13] and cation vacancy ($V_{Ga}$ in case of GaN). Among these defects, $O_N$ is the unavoidable impurity in group III-V nitrides which makes the GaN as an *n*-type semiconductor.[11,14] In the past, several attempts were made to investigate the effect of native defects on various physical properties like stoichiometry, morphology, growth rate, and luminescence efficiency in GaN nanowires (NWs).[15-20] In our previous report, we have studied the effect of native defects on controlling the carrier concentration and mobility.[21] Spectroscopic techniques such as Raman, photoluminescence (PL) were utilized to understand the electron-phonon coupling as well. The study demonstrated an increase in the electron-phonon coupling strength with an increase in the carrier concentration arising from native defects. However, these studies could hardly reveal any information on the local structure. X-ray photoelectron spectroscopy (XPS),[22] which helps in quantifying different chemical species in a material also fails in accurately quantifying the amount of $O_N$ defects in GaN. XPS is a surface-specific technique and gives information from the surface and sub-surface layers. In the case of GaN NWs, since the amount of surface adsorbed oxygen is dominant, XPS overestimates the oxygen impurity.[21] The microscopic distortion and local atomic configuration are not well understood through the above mentioned conventional spectroscopic techniques. Hence, it is pertinent to use a technique which can help in quantifying the defect content in GaN precisely. In this regards, X-ray absorption spectroscopy (XAS) is a suitable technique. XAS, being a local averaging technique, it helps in understanding the bond length, co-ordination number and symmetry around a core metal atom, which in turn governs the macroscopic properties of the material.[23] Therefore, XAS will act as a perfect tool to locate the site occupancy, determine the site symmetry of the defect and estimate the defects quantitatively.[24] XAS comprises of three regions *viz.* edge, pre-edge and post-edge. The edge denotes the sharp rise in absorption co-efficient at a particular energy specific to the orbital and the element being studied. While, 30 eV energy before the edge is known as pre-edge region. The pre-edge region, edge and 30 eV just after the edge region altogether is known as X-ray absorption near edge spectra (XANES). XANES region provides the information about local symmetry and various electronic transitions. 100-200 eV after the edge region is known as extended X-ray absorption fine structure (EXAFS). This region contains the information about radial distance and co-ordination number of the absorbing atom. In the recent past XAS is demonstrated as a suitable technique to understand many physical and chemical aspect of the materials. Faye *et al.*[25] determined the Eu oxidation state in the Eu implanted GaN and the related luminescence property using XAS analysis. Moreover, the analysis also clearly reavealed the Eu local environment in GaN NW for different fluence of Eu implantation. XAS also played a important role in understanding the core/shell nanostructure. Boldt *et al.*[26]

calculated the average coordination number of the probed element in the ZnSe/CdS nanocrystals with graded shells. The study provides information on interfacial alloying, cation migration, and atomic ordering at different growth temperature of ZnSe/CdS nanocrystals using EXAFS analysis. XAS is demonstrated in playing an important role in characterising the doping in a system. Negishi et al.[27] has determined the occupation site of Pd dopant in $Au_{24}Pd_1(SC_{12}H_{25})_{18}$ by investigating the Pd edge in EXAFS spectroscopy. The study suggest that the Pd atom preferentially goes to the center of $Au_{24}Pd_1(SC_{12}H_{25})_{18}$ to form the superatomic $Pd@Au_{12}$ core.

Another complementary technique to XAS is electron energy loss spectroscopy (EELS). However, EELS has its drawback as it requires special sample preparation. In addition, there are a scant number of reports on the studies of defects in GaN utilizing EELS. Fall *et al.*[28] have studied the line defects in GaN by utilizing EELS. However, it does not throw any light on the point defects. Patsha *et al.*[17] has qualitatively investigated the presence of point defect cluster in GaN NWs using EELS. However, the identification of point defect clusters using EELS analysis is rather speculative as supported by *ab initio* calculations and does not provide any information on the local atomic configuration viz. co-ordination number or bond length. Effect of point defects on local co-ordination and bond length decides the local crystal potential, which in turn decides the density of states of the material and subsequently control the macroscopic optoelectronic properties. Hence, XAS acts an excellent tool to determine the local atomic configuration and it has an obvious advantage over EELS concerning sample handling, as it does not require any special sample preparation and studies also can be performed on as-synthesized samples in ambience. To the best of our knowledge, we have identified the defect location in GaN for the first time by using EXAFS analysis and subsequently compared the experimental results with the insights from semi-empirical and density functional theory (DFT) calculations.

Similarly, in the context of alloying, the understanding of local structure is also of great importance. As alloying leads to the bandgap tailoring by modifying the local crystal potential, the information regarding the bond length and bond angle at the localized structure help in understanding the macroscopic properties.[5, 23] In the recent past, structural investigations of $CuFe_2O_4$ nanoparticles and $CuFe_2O_4-MO_2$ (M = Sn, Ce) nanocomposites by means of EXAFS analysis revealed the site occupancy of Cu and Fe copper in the matrix. The XAFS analyses of $CuFe_2O_4-SnO_2$ and $CuFe_2O_4-CeO_2$ nanocomposites show that the incorporation of the tetravalent metal ions does not change the local structure around Cu and Fe in $CuFe_2O_4$ nanoparticles.[29] Cosidering the above mentioned utilities and advantages of XAS technique, in this present study, we intend to investigate the local atomic configuration of nanostructured $Al_xGa_{1-x}N$ random alloy with the help of EXAFS analysis. The material of interest *i.e.* $Al_xGa_{1-x}N$ NW is synthesized *via* a novel technique of ion beam processing. In our previous study, we have demonstrated the synthesis of $Al_xGa_{1-x}N$ *via* ion beam techniques using GaN NW as the starting material.[30] The optical properties were studied using Raman and PL spectroscopies. Al atomic percentage (at.%) was calculated with the help of Raman spectroscopic analysis, which is indeed an indirect technique to calculate the alloy composition by exploiting Vegard's law.[30] Therefore, the actual information of the local structure and Al incorporation percentage is not clear from the previous study. In the present study, we intend to verify the Al incorporation at the Ga sites and other related information about the local structure by means of EXAFS analysis. In addition, EXAFS study is also utilized as a localized probe to determine the change in the bond length and co-ordination of Al atom with the neighboring atoms.

## 2. EXPERIMENTAL

### 2.1 Synthesis of GaN nanowires

GaN NWs with different III/V ratios were synthesized by the atmospheric pressure chemical vapor deposition (APCVD) technique by changing the carrier gas flow rate. The detailed growth process is reported in our previous article.[21] In brief, GaN NWs were synthesized on Si substrate by using Ga metal and $NH_3$ gas as nitrogen precursor and Ar as the carrier gas. Three samples named S1, S2, and S3 were synthesized with different flow rates of $NH_3$ and Ar. Sample S1 was grown under nitrogen rich conditions by flowing 50 sccm of $NH_3$ without any flow of Ar, while a mixture of precursor and carrier gases were utilized for the growth of sample S2 ($NH_3$ = 10 sccm + Ar = 10 sccm) and S3 ($NH_3$ = 10 sccm + Ar = 20 sccm).

### 2.2 Synthesis of $Al_xGa_{1-x}N$ nanowires

$Al_xGa_{1-x}N$ NWs were synthesized *via* ion beam techniques, namely ion beam mixing (IBM) and post-irradiation diffusion (PID) process using APCVD grown GaN NWs as the starting material. The detailed growth process is reported elsewhere.[30] In brief, in the IBM process, Al coated (~25 nm) as-grown GaN NW was irradiated with $Ar^+$ (25 keV) at fluences of 1E16 and 5E16 ions·$cm^{-2}$. In the PID process, prior to the coating of Al (~25 nm), the GaN NWs were irradiated with $Ar^+$ ion of the same energy and fluences as in the case of the IBM process. Both the processes were followed by annealing at 1000 ºC in $N_2$ (5N pure) atmosphere for 5 min.

### 2.3 Raman spectroscopic measurement

The vibrational studies were performed using Raman spectroscopy (inVia, Renishaw, UK) with laser excitation of 514.5 nm. The scattered photons were dispersed through 1800 groves·$mm^{-1}$ grating to the CCD detector. The resonance Raman spectra were collected with the 325 nm laser excitation and the scattered photons were dispersed through a grating of 2400 groves·$mm^{-1}$.

## 2.4 X-ray absorption spectroscopic measurement

XAS studies were carried out by measuring Ga *K*-edge absorption spectra (Fig S1, Supporting Information) in fluorescence mode at the beamline of the INDUS-2 synchrotron source (2.5 GeV, 200 mA) at the Raja Ramanna Centre for Advanced Technology (RRCAT), Indore, India.[31] The beamline operates in the energy range of 4 to 25 keV with energy resolution *E*/d*E*~10. The beamline optics consist of an Rh/Pt coated collimating meridional cylindrical mirror followed by a Si (111) ($2d = 6.2709$ Å) based double crystal monochromator (DCM). The second crystal of the DCM is a sagittal cylindrical crystal used for horizontal focusing. One ionization chamber (with a length of 300 mm) and a solid-state detector were used. The photon energy scale was calibrated by measuring the XAS spectrum of a standard Ga metal foil. The data was acquired in fluorescence mode with standard 45° geometry. The absorption coefficient $\mu$ was obtained using the relation: $\mu(E) \propto I_0/I_f$; where $I_0$ is the incident X-ray intensity, and $I_f$ is the measured fluorescence intensity.

For the EXAFS analysis, the oscillation part of the measured absorption co-efficient ($\mu(E)$) was converted to the fine structure-function ($\chi(E)$) using the following equation, taking into account the threshold energy ($E_0$), edge step ($\Delta\mu_0(E_0)$) and the smooth background function as the absorption of an isolated atom represented by $\mu_0(E)$:

$$\chi(E) = \frac{\mu(E) - \mu_0(E)}{\Delta\mu_0(E_0)} \quad \ldots\ldots\ldots\ldots\ldots\ldots\ldots\ldots\ldots\ldots\ldots\ldots\ldots (1)$$

Analysis of EXAFS is performed in *k*-space by converting the absorption co-efficient $\chi(E)$ to *k*-space ($\chi(k)$) using the following relation:

$$k = \sqrt{\frac{2m(E-E_0)}{\hbar^2}} \quad \ldots\ldots\ldots\ldots\ldots\ldots\ldots\ldots\ldots\ldots\ldots\ldots\ldots (2)$$

where *m* is the mass of the electron, $\chi(k)$ is weighted by $k^3$ to amplify oscillations at high *k*. The $k^3\chi(k)$ functions were Fourier transformed to *R*-space using the ARTEMIS software package. χ(R) versus R (radial distances) plots were then used to derive the co-ordination number and radial distances of the atoms in various co-ordination spheres.

For the fitting of the spectra, crystallographic inputs of the wurtzite phase of GaN was considered. The real space fittings were performed between 1 to 4 Å. The Fourier transform was carried out with multiple $k$-weight. It is known that reduced $\chi^2$ is far greater than 1 in the case of EXAFS due to inadequate assessment of the uncertainty in data and imperfections in the calculations that go into the model. Hence, the quality of the fit was measured by minimizing an alternate statistical parameter; known as $R$-factor. $R$-Factor is defined by the following equation[31]

$$R = \frac{\sum_{i=min}^{max}[Re(\chi_d(r_i)-\chi_t(r_i))^2 + Im(\chi_d(r_i)-\chi_t(r_i))^2]}{\sum_{i=min}^{max}[Re(\chi_d(r_i))^2 + Im(\chi_d(r_i))^2]} \quad\ldots\ldots\ldots\ldots\ldots\ldots (3)$$

Four path parameters ($S_0^2$= amplitude reduction factor, $\Delta r$ = change in half path length, $\Delta E_0$ = energy shift, $\sigma^2$= mean square displacement) for each path were taken into consideration for the fitting of the experimental data with FEFF generated modeling using ARTEMIS, which had IFEFFIT integration. Only the single scattering paths were included in the analysis and the multiple scattering paths were discarded because of low rank. The number of free variables was always kept below the upper limit set by the Nyquist theorem ($N_{free}$ =2$\Delta k\Delta r/\pi$+ 1), by assuming the same $E_0$ for all the paths. The overall value of the $R$-factor was then minimized to establish the quality of the fitting. The details about the path for GaN and Al$_x$Ga$_{1-x}$N is presented in their respective result and analysis section.

## 2.5 Density functional theory and semi-empirical calculation

For the DFT calculations, we used the QUANTUM ESPRESSO software[32] to investigate the structural changes on the GaN bulk doped with oxygen atoms and Ga vacancies. In our calculations, we considered a GaN supercell system composed of 48 atoms. The calculations were performed within the generalized gradient approximation (GGA) using the PBE functional[33] to describe the exchange-correlation term. The valence electrons were described by a plane-wave basis set with cut-off energy of 50 Ry, and a Γ- centered Monkhorst-Pack[34] 4x4x5 k- mesh was used to sample the Brillouin Zone. These parameters were chosen for reproducing the lattice parameter value close to the reference [35]. We also carried out an additional geometrical optimization analysis using the semi-empirical MOPAC2016 software.[36] The results obtained with MOPAC-PM6-DH2 parameterization produced the same DFT trends.

## 3. RESULT AND DISCUSSION

### 3.1 Compositional Analysis of GaN

Three sets of GaN NWs named S1, S2, and S3 were synthesized with different flow rates of the precursor gas in the APCVD technique. The detailed compositional analysis of these GaN NWs samples is estimated from the XPS analyses, and their constituent elemental compositions are presented in Table 1.

**Table 1:** The atomic percentage of constituent elements in GaN NWs estimated from the XPS analyses

| Sample | Ga (at. %) | N (at. %) | O (at. %) |
| --- | --- | --- | --- |
| S1 | 29.9(1) | 41.8(1) | 28.3(1) |
| S2 | 27.9(1) | 40.1(1) | 32.0(1) |
| S3 | 26.7(1) | 39.9(1) | 33.4(1) |

The presence of the oxygen atoms, as revealed from the XPS study resulted from their unintentional incorporation in GaN during the APCVD growth process. It is observed that oxygen percentage increases from 28.3% in S1 to 32 and 33.4 % in S2 and S3, respectively. It is to be noted that in S1 the nitrogen precursor was 50 sccm, however in S2 and S3 the nitrogen precursor flow was maintained at 10 sccm and only the carrier gas concentration was varied. It is known that Ga has a higher affinity towards oxygen over nitrogen. Hence, in a nitrogen depleted condition oxygen incorporation happens though it is unwanted. The source of this oxygen can be attributed to the low base vacuum of the chamber and the insufficient amount of precursor gases (NH$_3$) to reduce the presence of oxygen at the growth time. In addition to this, the substrate and quartz tube used for the growth can also act as a source of oxygen. Apart from the oxygen related defects in the sample, presence of the SiO$_x$ layer on Si, which was used as a substrate to grow the GaN NWs, might have also contributed to the observed atomic percentage of oxygen in all the samples. However, as the SiO$_x$ layer is common in all samples the relative variation in oxygen concentration from sample S1 to S3 can be attributed to the oxygen impurities from GaN lattice.

From the XPS analysis, it is observed that Ga and nitrogen concentration decrease and oxygen concentration increases from S1 to S3. The observed increase in oxygen concentration from sample S1 to S3 is due to the reduction in the nitrogen concentration during the growth. It may be emphasized that since S3 is grown under less concentration of NH$_3$, the amount of oxygen involved in the growth process is expected to be high, by forming O$_N$ point defects.

### 3.2 Raman spectroscopic analysis of GaN nanowires

The Raman spectra of NWs from all of the GaN samples are represented in Figure 1. In case of S1, the peaks centered at ~567, and 721 cm$^{-1}$ correspond to the Raman active $E_2$(high) phonon mode, and longitudinal optical (LO) phonon modes of $A_1$ symmetry ($A_1$(LO)), respectively, thus confirming the presence of a wurtzite GaN phase.[37,38] Along with the Raman active modes of wurtzite structure, a broad peak centered ~628 cm$^{-1}$ is assigned to the surface optical (SO) phonon modes of GaN corresponding to $A_1$ symmetries.[39] The peak centered at ~521 cm$^{-1}$ corresponds to the Si substrate on which GaN NWs were grown (Figure 1a). The Raman spectra of GaN NWs of sample S2 (Figure 1b) also showed similar phonon modes to that of S1. However, $A_1$(LO) mode is blue-shifted in S2 with 3 cm$^{-1}$ as compared to the sample S1. In the case of sample S3 (Figure 1c), $A_1$(LO) mode is further blue-shifted by 6 cm$^{-1}$ as compared to sample S1. It is noteworthy to mention that the blue-shift is observed only in the case of $A_1$(LO) phonon mode, while all other phonon modes do not show any significant shift across the samples.

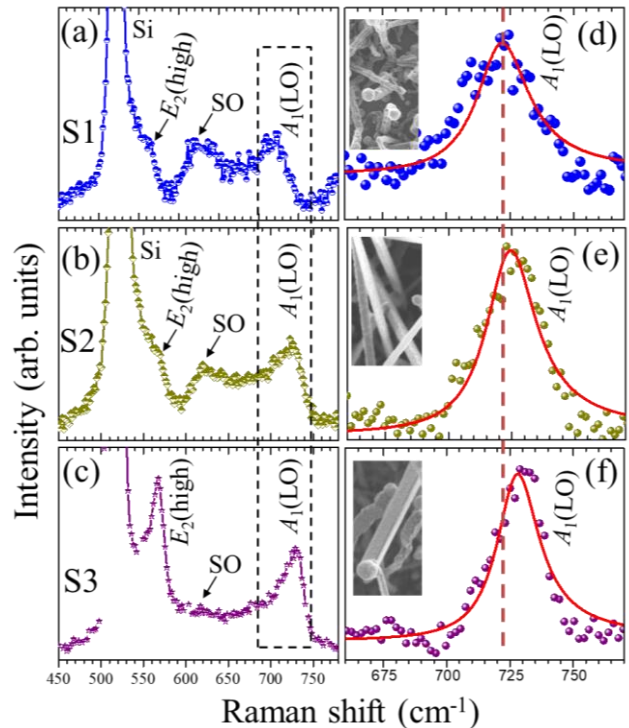

**Figure 1.** Typical Raman spectra of GaN NWs sample (a) S1, (b) S2, and (c) S3. The corresponding fitted spectra of $A_1$(LO) phonon mode is represented in (d)-(f), with the scanning electron micrograph as inset. The dotted rectangular box and dotted lines are guide to an eye for the observed blue-shift of $A_1$(LO) phonon mode as one move from sample S1 to S3.

The observed blue-shift in the $A_1$(LO) phonon mode of NWs from samples S1 to S3 (Figure 1d-1f) could be due to the interactions among the LO phonons and free charge carriers in the samples, as such a shift is not possible in $A_1$(LO) phonons alone if it is associated with strain or temperature effect. As revealed by the XPS analysis, the GaN NWs are expected to possess different carrier concentrations as a function of native defects present in the system. The defect induced free charge carriers thus can strongly interact with LO phonons, resulting in the blue-shift of $A_1$(LO) phonon modes from samples S1 to S3. $A_1$(LO) mode is a polar mode and can be influenced by the free carriers in the system. The $A_1$(LO) mode blue-shifted by ~3 cm$^{-1}$ in case of sample S2 and 6 cm$^{-1}$ in S3 as compared to S1 is indicative of the fact that the carrier concentration in sample S2 and S3 are higher as compared to S1. The carrier concentration in the sample is a function of doping concentration and defect incorporation. It is well known that the native defects in III-V nitrides influence the carrier concentration.[5] The signature of carrier concentration in $A_1$(LO) modes is indicative of the presence of defects in the system. From the Raman line shape analysis, we have calculated the carrier concentration and mobility in these samples and reported in the previous article.[21] For the convenience of the readers, the values are provided in Table T1 (Supporting Information). The carrier concentration increases as one moves from sample S1 to S3 with a decrease in mobility. Since the NWs are not extrinsically doped, the increase in carrier concentration implies the presence of donor type native defects in the sample. Moreover, the defect density increases from sample S1 to S3. However, the complete scenario of the defect incorporation in the local structure of the GaN is not clear from the above spectroscopic technique. The effect of defects on local lattice structure and quantification of such defects is performed in the following section utilizing EXAFS analysis.

**3.3 Local structure and atomistic studies of oxygen doping in GaN nanowires**

XAS studies were performed for all samples at the Ga $K$-edge. For the fitting of GaN XAS spectra, three paths corresponding to three basal nitrogen's [(Ga–N)bs] and one path corresponding to the axial nitrogen [(Ga–N)ax] from the first co-ordination sphere and one path from the second co-ordination sphere (Ga–Ga) were considered. The overall value of $R$-factor was then minimized to establish the quality of the above fitting. The fitted EXAFS oscillation χ(k) curves in $R$-space are shown in Figure 2. The bond lengths of Ga–N and Ga–Ga paths as well as the corresponding co-ordination numbers (CN) of each path ($N \cdot S_0^2$, $N$ = degeneracy of path), Debye-Waller factor ($\sigma^2$), and $R$-factor for the samples are tabulated in Table 2.

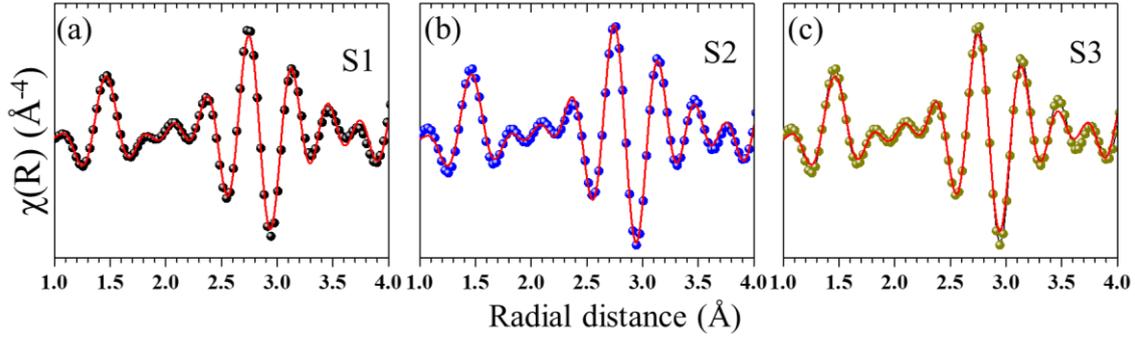

**Figure 2.** The extended region fitting of GaN NWs sample (a) S1, (b) S2 and (c) S3 in *R*-space.

It is known that oxygen and nitrogen are indistinguishable scatterers when one is analyzing Ga *K*-edge spectra.[40] To make it distinguishable, two methods were tried. One is to replace the nitrogen with oxygen in GaN, and FEFF calculation is performed to find out if it yields any paths which are distinguishable from Ga-N paths. However, such an exercise did not yield any extra paths for Ga-O. In another method, an extra Ga-O path from gallium oxide crystal structure inputs was generated and was tried for fitting, which resulted in a very high R-factor. Hence, in the present analysis, though the nitrogen and oxygen are indistinguishable, the resulting change in bond length can still be safely attributed to oxygen incorporation in conjunction with the concentration of oxygen as observed from XPS.

**Table 2.** Fitted path parameters for the GaN NW with different III/V ratios.

| Sample | Path | $N \cdot S_0^2$ CN | $R_{eff} + \Delta R$ Bond length (Å) | $\sigma^2$ | R factor |
|---|---|---|---|---|---|
| S1 | Ga-N1 | 0.77(1) | 1.90(1) | 0.005(1) | |
| | Ga-N2 | 3.07(3) | 1.93(0) | 0.001(2) | 0.03 |
| | Ga-Ga | 8.51(2) | 3.16(3) | 0.002(1) | |
| S2 | Ga-N1 | 1.20(3) | 1.91(0) | 0.005(1) | |
| | Ga-N2 | 2.87(2) | 1.93(2) | 0.001(1) | 0.02 |
| | Ga-Ga | 13.40(2) | 3.17(2) | 0.005(1) | |
| S3 | Ga-N1 | 0.90(2) | 1.65(0) | 0.001(0) | |
| | Ga-N2 | 2.93(1) | 1.92(1) | 0.003(1) | 0.02 |
| | Ga-Ga | 14.60(3) | 3.17(3) | 0.007(2) | |

For the GaN samples of S1, S2, and S3, the co-ordination numbers of Ga are found to be 3.84, 4.04, 3.83, respectively, in the first co-ordination sphere. In the tetrahedral structure, Ga has a co-ordination number of four with three basal nitrogen and one axial nitrogen.[5, 41] The bond length of the axial Ga-N in the axial direction is found to be 1.9, 1.9, 1.65 for S1, S2, and S3, respectively. The compression in axial bond length for S3 can be because of the higher amount of oxygen in sample S3 as compared to that of S1 and S2.

In order to shed light on the observed result, we have performed density functional theory (DFT) calculation on the optimized wurtzite GaN structure. We obtained the lattice parameter value as $a = 3.22$ Å, similar to that of the earlier reported study with the same DFT parameters.[35] In Figure 3, one can see there are three basal nitrogen and one axial nitrogen bonded with Ga atom. The three basal nitrogen atoms are denoted as 1, 2 and 3, and one axial nitrogen as 4. The bonds between basal nitrogen and Ga atom are found to be 1.97 Å ($b_1 = b_2 = b_3$) and the bond with axial nitrogen is 1.98 Å ($b_4$).

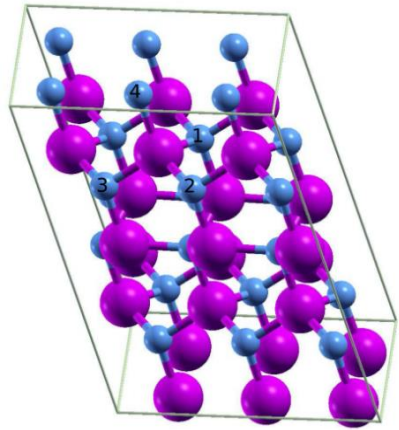

**Figure 3.** The GaN configuration used in the DFT calculations showing the basal and axial nitrogen atoms

The difference between basal and axial Ga-N bond length occurs only due to numerical precision (rounding approximation). To understand the nature of Ga-N and Ga-O bond lengths when dopant oxygen replaces various nitrogen position, we replaced each nitrogen atom in the tetrahedron one by one (configuration shown in Figure 3), The calculated bond lengths for Ga-N and Ga- O are presented in Table 3. It is observed from Table 3 that the Ga-N bond length in the oxygen substituted GaN tetrahedrons remains the same as that of unsubstituted Ga-N. However, Ga-O bond elongates as compared to that of Ga-N bonds in all the substituted configurations, regardless of the basal or axial position

**Table** 3. Calculated Ga-N and Ga-O bond length in basal and axial positions

| Configuration | $b_1$ (Å) | $b_2$ (Å) | $b_3$ (Å) | $b_4$ (Å) |
|---|---|---|---|---|
| GaN | 1.97(2) | 1.97(2) | 1.97(2) | 1.98(3) |
| GaN ($O^1$) | 2.05(6) | 1.97(2) | 1.96(2) | 1.96(3) |
| GaN ($O^2$) | 1.97(2) | 2.05(6) | 1.96(2) | 1.96(3) |
| GaN ($O^3$) | 1.96(2) | 1.97(2) | 2.05(6) | 1.96(3) |
| GaN ($O^4$) | 1.95(1) | 1.95(1) | 1.95(1) | 2.07(8) |

In the EXAFS study (Table 2), we see a reduction in the axial bond length with the incorporation of oxygen impurity, which is in contradiction with the DFT calculation. From the elemental analysis performed using XPS (Table 1), one can observe that the Ga elemental percentage decreases with increase in the oxygen incorporation, as we move from sample S1 to S3. Therefore, it is pertinent to consider the presence of $V_{Ga}$ while calculating the bond length using DFT analysis. In the theoretical and experimental studies,[5,17] it is well reported that the $V_{Ga}$ and $V_{Ga}$-$O_N$ ($V_{Ga}$ bonded to $O_N$) complex are present in the GaN system with the incorporation of oxygen. The oxygen atom incorporated during the growth process, enhance the formation of $V_{Ga}$ by forming energetically stable vacancy-impurity ($V_{Ga}$-$O_N$) complexes. Therefore, to emulate the experimental condition and to enrich our understanding, the presence of $V_{Ga}$ in the system was considered while performing DFT calculation.

In an earlier study,[42] it is reported that the axial and basal bond length increase for both pristine and oxygen incorporated ($O_N$) system for one $V_{Ga}$ per unit cell. This is because $V_{Ga}$-$O_N$ bond length is greater than that of Ga-$O_N$ bond due to the electrostatic interactions. When $V_{Ga}$ interacts with $O_N$, oxygen atom moves away from the Ga vacancy because of small electrostatic repulsion. In the present study, we have relaxed structures (from DFT calculation) by considering the interaction of one substituted oxygen with two Ga vacancy ($2V_{Ga}+ O_N$) and two substituted oxygen from two neighboring GaN tetrahedrons with two Ga vacancy ($2V_{Ga}+2 O_N$) as shown in Figures 4a and 4b, respectively. In these cases, we observe that $V_{Ga}$ produces small electrostatic repulsion to $O_N$, and as a consequence, the oxygen atom is pushed towards the Ga-$O_N$ axial bond, causing the bond length to compress.

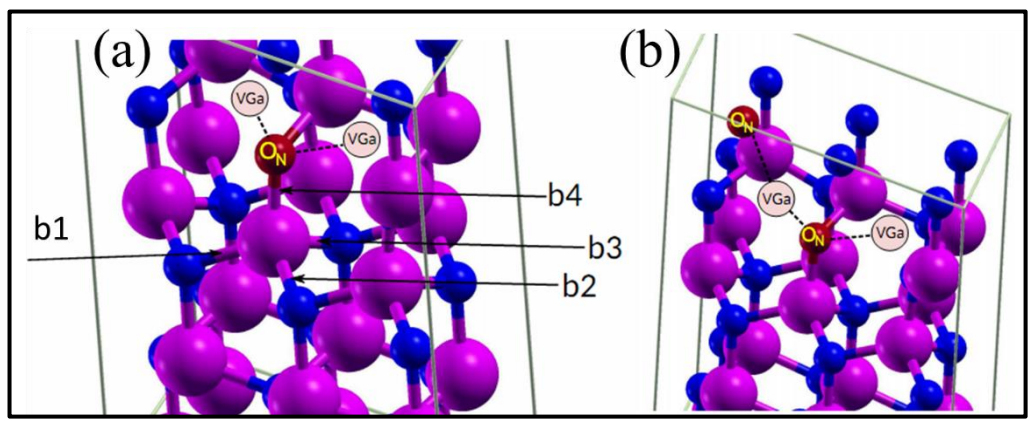

**Figure 4.** The configuration of GaN (a) $2V_{Ga}+ O_N$ and (b) $2V_{Ga}+ 2O_N$ used for DFT calculations

Table 4 shows the results for axial and basal bonds for the configurations described above (Figure 4). The axial bond length in all cases is contracted. As the bond distance, derived from EXAFS analysis, is averaged over many tetrahedrons and DFT calculations become very costly for larger systems. A semi-empirical method (PM6-DH2) was used to calculate bond distances for three neighboring tetrahedrons. Table 4 shows the bond distances calculated by using PM6-DH2 along with the DFT calculation. In the present study, the configurations investigated in Tables 3 and 4, were proposed and discussed in the literature on experimental and theoretical works.[43-52] In the cases of high oxygen concentration, it is favorable for the oxygen to occupy the $V_N$ and to bind to $V_{Ga}$, which is randomly distributed within the bulk.[50]. In some cases, it was observed that $V_{Ga}$ can be distributed by a line point defect.[52] Subsequently, with the introduction of oxygen in the GaN bulk, it is energetically favorable for $V_{Ga}$ and oxygen atoms to bind. Some studies investigated also the interaction between $V_{Ga}$ pairs binding to oxygen pairs.[44] The interaction, in this case, tends to be attractive and there may be diffusion of these defects,[51] which enables to generate configurations with some $V_{Ga}$ very close to itself.[50]

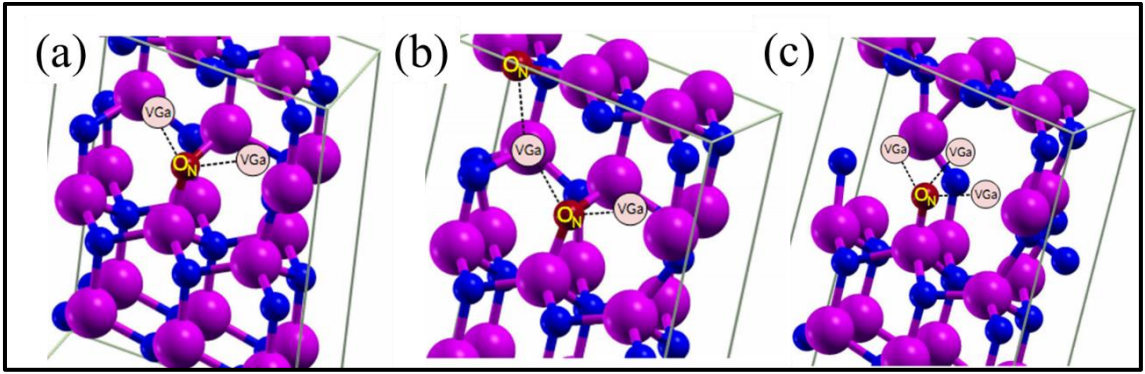

**Figure 5.** The configuration of GaN (a) $2V_{Ga}+ O_N$, (b) $2V_{Ga}+ 2O_N$ and (c) $3V_{Ga}+ O_N$ used for PM6-DH2 calculation.

One can observe that DFT and Pm6-DH2 results are similar for configuration $2V_{Ga}+O_N$ and $2V_{Ga}+2O_N$. For $3V_{Ga}$ configuration, the DFT method does not converge with the same parameters, therefore we have used the semi-empirical method Pm6-DH2 in order to obtain the converged geometry showed in figure 5. It is observed from Table 4 that for $3V_{Ga}+O_N$ configuration, the value for the axial bond length is 1.66 Å, which is very close to the experimental results for sample S3 (1.65 Å). This observation implies that there is no distortion in the tetrahedral in the basal site. Hence, one can conclude that the oxygen incorporation as dopant at nirogen site in GaN structure is favorable in the axial position rather than the basal position.

**Table 4**. Calculated Ga-N and Ga-O bond length in basal and axial positions in the configuration mentioned in Figure 4 and 5

| Method | Configuration | Figure | $b_1$ (Å) | $b_2$ (Å) | $b_3$ (Å) | $b_4$ (Å) |
|---|---|---|---|---|---|---|
| DFT | $2V_{Ga}+O_N$ | Figure 4a | 1.99(4) | 1.99(4) | 1.99(4) | 1.84(10) |
| DFT | $2V_{Ga}+2O_N$ | Figure 4b | 1.99(4) | 1.99(4) | 1.98(3) | 1.85(11) |
| PM6-DH2 | $2V_{Ga}+O_N$ | Figure 5a | 2.04(6) | 2.09(9) | 2.11(10) | 1.84(10) |
| PM6-DH2 | $2V_{Ga}+2O_N$ | Figure 5b | 2.04(6) | 2.08(8) | 2.11(10) | 1.83(10) |
| PM6-DH2 | $3V_{Ga}+1O_N$ | Figure 5c | 2.18(13) | 2.29(18) | 2.21(15) | 1.66(1) |
| **Experimental** | **Sample S3** | ---- | **1.92(1)** | **1.92(1)** | **1.92(1)** | **1.65(0)** |

### 3.4 Raman spectroscopic analysis of $Al_xGa_{1-x}N$

The typical resonance Raman spectra of the samples irradiated with a fluence of 1E16 ions·cm$^{-2}$ and 5E16 ions·cm$^{-2}$ in the IBM and the PID process are shown in Figure 6. Along with the first order $A_1$(LO) mode, the second order $2A_1$(LO) mode is also observed in the as-grown sample. In the case of the IBM and PID samples, there is a clear blue-shift of $A_1$(LO) mode, which suggests the one mode phonon behavior in the random alloy model. The Al incorporation percentage is calculated using the band bowing formalism of the random alloy model. The atomic percentage of Al in $Al_xGa_{1-x}N$ random alloy increases with an increase in the irradiation fluence and post irradiation annealing temperature in both IBM and PID process. A detailed analysis is reported in our previous study.[30] For the reader's convenience, Al at.% in tabulated form is provided in the Supporting Information (Table T2). However, localized information regarding alloy formation is not adequate. Therefore, the EXAFS analysis is performed in the following section to get a clear picture of the localized information at the atomic scale.

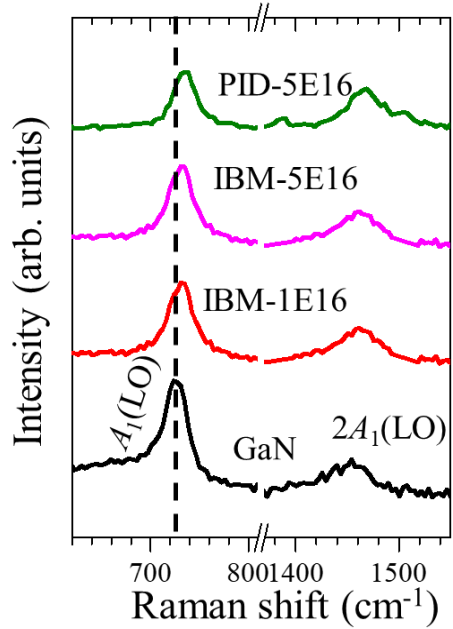

**Figure 6.** Typical Raman spectra of GaN NWs and $Al_xGa_{1-x}N$ NWs synthesized in different conditions.

### 3.5 Local structure studies of $Al_xGa_{1-x}N$ nanowire

For the fitting, the EXAFS spectra of $Al_xGa_{1-x}N$, one Ga was replaced with an Al in the first co-ordination sphere of GaN in the FEFF calculations, which resulted in single scattered paths of only Ga-N paths in the first co-ordination sphere and Al was seen to replace Ga only in the second co-ordination sphere. A separate run in FEFF was carried out to investigate if the Al generates any other paths when it completely replaces Ga as its core. The study was performed by using similar crystallographic inputs of GaN by changing the core to Al. The resulting first co-ordination sphere paths had similar path lengths as that of Ga-N paths wherein only a small percentage of Al was replaced initially in the FEFF calculations. Hence, the former calculations were utilized for the fitting of $Al_xGa_{1-x}N$ spectra. Al in GaN was accounted for the fitting by replacing one Al in place of Ga at a distance of 3.17 Å in FEFF calculations. The calculation generated two kinds of paths *i.e.* Ga-Al and Ga-Ga at a distance of 3.18 Å. As the co-ordination number is $N \cdot S_0^2$ and ideally the second co-ordination sphere has 12 Ga atoms, the concentration of Al was accounted for by parametrizing $S_0^2$ of Ga-Al as 'y' and that of Ga-Ga as 12-y. The fitted EXAFS oscillation χ(k) curves in *R*-space are shown in Figure 7 for all the $Al_xGa_{1-x}N$ samples.

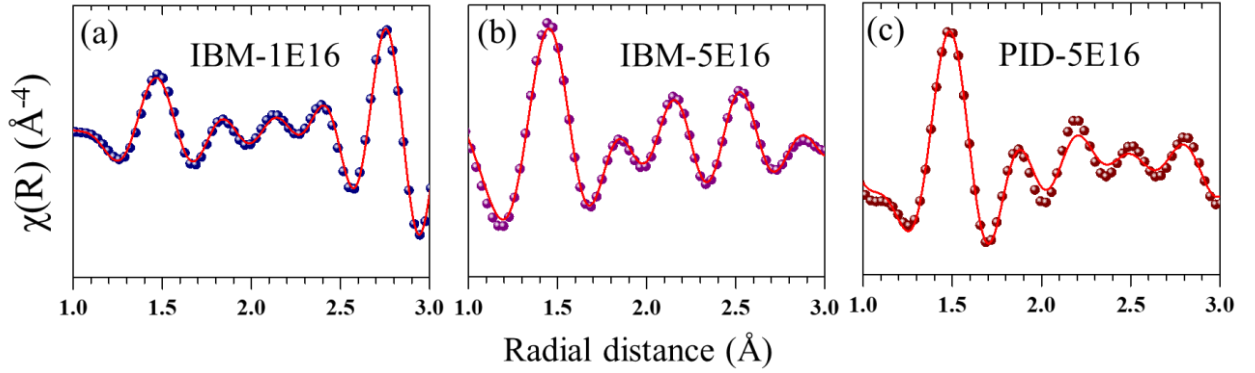

**Figure 7.** The extended region fitting of $Al_xGa_{1-x}N$ NWs sample (a) synthesized in an IBM process with an irradiation fluence of 1E16 ion·cm$^{-2}$ and (b) 1E16 ion·cm$^{-2}$ (c) synthesized in a PID process with an irradiation fluence of 5E16 ion·cm$^{-2}$

The bond lengths of Ga-N and Ga-Ga paths, the corresponding CN of each path, $\sigma^2$ and R-factor values are tabulated in Table 5. In all the samples Al replaces Ga in the GaN matrix irrespective of the synthesis methodology (IBM and PID), consistent with the study performed on $Al_xGa_{1-x}N$ films.[53] It is observed from the co-ordination number of Ga-Al and Ga-Ga paths (Table 5) that the Ga-Al co-ordination number increases from 0.43 to 4.69 as the fluence increases from 1E16 to 5E16 ions·cm$^{-2}$ in the IBM process.

**Table 5.** Fitted path parameters for the $Al_xGa_{1-x}N$ NW synthesized at different conditions with a variation of Al content

| Sample | Path | $N \cdot S_0^2$ CN | $R_{eff}+\Delta R$ Bond length (Å) | $\sigma^2$ | R-factor |
|---|---|---|---|---|---|
| AlGaN-IBM-1E16 | Ga-N1 | 1.2(1) | 1.54(1) | 0.019(1) | 0.04 |
|  | Ga-N2 | 3.6(3) | 1.91(4) | 0.001(2) |  |
|  | Ga-Al | 0.43(3) | 3.12(3) | 0.005(0) |  |
|  | Ga-Ga | 11.56(2) | 3.18(1) | 0.005(3) |  |
| AlGaN-IBM-5E16 | Ga-N1 | 1.2(3) | 1.80(2) | 0.003(3) | 0.02 |
|  | Ga-N2 | 3.6(2) | 1.91(4) | 0.003(2) |  |
|  | Ga-Al | 4.69(3) | 2.96(3) | 0.014(1) |  |
|  | Ga-Ga | 7.30(2) | 2.93(1) | 0.013(2) |  |
| AlGaN-PID-5E16 | Ga-N1 | 1.20(4) | 1.79(1) | 0.003(1) | 0.03 |
|  | Ga-N2 | 2.92(4) | 1.96(2) | 0.004(2) |  |
|  | Ga-Al | 2.69(3) | 3.12(3) | 0.005(1) |  |
|  | Ga-Ga | 9.30(4) | 3.08(2) | 0.012(2) |  |

The observation is consistent with the Raman spectroscopic analysis performed in our previous study.[30] The co-ordination number of Ga-Al in PID process at a fluence of 5E16 ions·cm$^{-2}$, however, yields only a value of 2.69. Hence, it can be concluded that the IBM method has a distinct advantage over the PID one to incorporate Al in GaN, as it has the potential to incorporate a higher amount of Al with the same irradiation fluence (5E16 ions·cm$^{-2}$). It is also interesting to observe that the bond length of the axial Ga-N bond is shortened in the case of all Al$_x$Ga$_{1-x}$N samples (Table 5) as compared to as-grown GaN (Table 2). A detailed calculation may be needed to understand the Al$_x$Ga$_{1-x}$N system prepared by a novel ion beam techniques involving complex defect formation and their stabilization in the annealing process.

## 4. CONCLUSIONS

The local structure of GaN nanowires (NWs) with different native defect density and Al$_x$Ga$_{1-x}$N random alloy with different Al atomic percentage is studied by X-ray absorption spectroscopy and computer simulations. The co-ordination number and bond length of the local tetrahedral structure is determined from extended X-ray absorption fine structure analysis. In all the GaN NW samples, oxygen as dopant occupies the lattice position of nitrogen. The sample with a higher amount of dopant oxygen shows a compression in the axial bond length without any distortion of the basal bond length of Ga-N, which implies the favorable position of oxygen dopant as an axial site of the tetrahedral. In the case of Al incorporation to GaN NWs, the Al occupies the Ga position by reducing the Ga-N bond length. Moreover, the study also reveals the fact that the ion beam mixing process is a better choice as compared to post-irradiation diffusion process for the formation of Al$_x$Ga$_{1-x}$N random alloy. Thus, the study provides a clear understanding of macroscopic properties evident from the microscopic structure in III-V nitrides.

**ACKNOWLEDGEMENTS**

The authors sincerely thank A. K. Yadav, S. N. Jha and D. Bhattacharyya of AMPD, BARC, Mumbai for their help in EXAFS measurement and data fitting. We also thank R. Pandian of SND IGCAR and S. Bera of BARCF, Kalpakkam for their respective help in FESEM and XPS study. RMT and DSG thank the Center for Computing in Engineering and Sciences at Unicamp for financial support through the FAPESP/CEPID Grants #2013/08293-7 and #2018/11352-7 and the Brazilian Agencies CNPq and CAPES.

**SUPPORTING INFORMATION**

Ga-*K* edge EXAFS spectra of GaN and $Al_xGa_{1-x}N$ NW samples (Figure S1), carrier concentration and mobility of the GaN NWs calculated from Raman spectroscopic analysis (Table T1), Al atomic percentage in $Al_xGa_{1-x}N$ with different fluences and annealing temperatures for IBM and PID processes (Table T2) supplied as supporting information.